\documentclass[11pt]{article}
\usepackage{mathrsfs,graphicx,bm,amsmath,blois,epsfig}

\begin{document}

\bibliographystyle{unsrt}

\title{\bf Isocurvature Perturbations in Quintessence Cosmologies}

\author{Gregor Sch{\"{a}}fer}

\address{Institut f{\"{u}}r Theoretische Physik, Philosophenweg 16, 69120 Heidelberg, Germany}

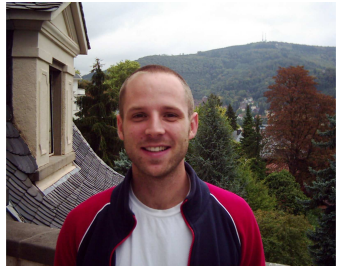
\newcommand{\norm}[1]{\mathcal{N}_{\rm #1}}

\def\dblone{\hbox{$1\hskip -1.2pt\vrule depth 0pt height 1.6ex width 0.7pt
            \vrule depth 0pt height 0.3pt width 0.12em$}} 
\newcommand{\Psiflex}{\Phi}
\newcommand{\Phiflex}{\Psi}

\newcommand{\isocdm}{CDM isocurvature}
\newcommand{\isobar}{baryon isocurvature}
\newcommand{\isonu}{neutrino isocurvature}
\newcommand{\Vphi}{V_{q}}
\newcommand{\Vphitilde}{\tilde{V}_{q}}
\newcommand{\Vphitildedot}{\dot{\tilde{V}}_{q}}
\newcommand{\Vphidot}{{\dot V}_{q}}
\newcommand{\Deltaphi}{\Delta_{q}}
\newcommand{\Deltaphidot}{{\dot\Delta}_{q}}
\newcommand{\wphi}{w_{q}}
\newcommand{\wq}{\wphi}
\newcommand{\Omegaphi}{\Omega_{q}}
\newcommand{\Mp}{{\rm M}_{\bar P}}

\newcommand{\Dg}[1]{\Delta_{#1}}
\newcommand{\rd}{\textrm{d}}

\newcommand{\hubble}{\mathcal{H}}

\newcommand{\dotphi}{\dot{\bar\varphi}}

\newcommand{\adota}{\frac{\dot a}{a}}

\newcommand{\ee}{\begin{equation}}
  \newcommand{\eee}{\end{equation}}
\newcommand{\ea}{\begin{eqnarray}}
  \newcommand{\eea}{\end{eqnarray}}

\newcommand{\ddotphi}{\ddot{\bar\varphi}}

\newcommand{\mpc}{\textrm{Mpc}}
\newcommand{\cpm}{\textrm{Mpc}\ensuremath{^{-1}}}
\newcommand{\cmbeasy}{CMBEASY\ }

\newcommand{\ome}[2][]{\ensuremath{
    \Omega^{#1}_{#2}}}
\newcommand{\omebar}[2][]{\ensuremath{
    \bar{\Omega}_{#1}^{#2}}}
\newcommand{\omc}{\ome[0]{c}}
\newcommand{\omb}{\ome[0]{b}}
\newcommand{\omq}{\ome[0]{\varphi}}
\newcommand{\omca}{\Omega_{c}}
\newcommand{\omba}{\Omega_{b}}
\newcommand{\omqa}{\Omega_{q}}
\newcommand{\omga}{\Omega_{\gamma}}
\newcommand{\omna}{\Omega_{\nu}}
\newcommand{\omma}{\Omega_{m}}

\maketitle\abstracts{
We present a systematic treatment of the initial conditions
and evolution of cosmological perturbations in a universe
containing photons, baryons, neutrinos, cold dark matter, and a
scalar quintessence field.
By formulating the evolution in terms of a differential equation involving a matrix acting on a vector
comprised of the perturbation variables, we can use the familiar language of eigenvalues and
eigenvectors. As the largest eigenvalue of the evolution matrix is fourfold degenerate, it follows
that there are four dominant modes with non-diverging gravitational potential at early times, corresponding to 
adiabatic, cold dark matter isocurvature, baryon isocurvature and neutrino isocurvature perturbations. We conclude
that quintessence does not lead to an additional independent mode.}


\subsection*{Introduction}

We present here a systematic treatment of initial conditions for 
quintessence models. The universe we will consider contains photons, baryons, neutrinos, 
cold dark matter and a scalar quintessence field. We will formulate the evolution equations for the perturbation
variables as a first order differential matrix equation:
\parbox{4.5cm}{\begin{equation}\label{eqn::matrix}
\frac{\textrm{d}}{\textrm{d} \,\ln x} \bm{U} = A(x) \bm{U},
\end{equation}} \;\; \;
where the vector $\bm{U}$ contains all perturbation variables and
the matrix $A(x)$ encodes the evolution equations.
In doing so,
we relate the problem of finding initial conditions and dominant modes
to the familiar language of eigenvalues and eigenvectors. This
formulation makes ``mode-accounting'' transparent by counting the
degeneracy of the largest eigenvalue.
We find four dominant modes that remain regular
at early times. For physical reasons, we choose a basis 
using 
adiabatic, \isocdm, \isobar\ and \isonu \ initial conditions.

\subsection*{The Different Modes}
In the following we  adopt
the  gauge-invariant approach as devised by Bardeen \cite{Bardeen:kt}. For a more detailed derivation of the perturbation
equations see Doran et al.\cite{Doran:2003xq}.
It turns out that the evolution is best described as a function of $x
\equiv k \tau$, where $\tau$ is the conformal time and $k$ the comoving
wavenumber of the mode. We assume that at early times, the universe expands as if radiation dominated. Assuming 
tracking quintessence we obtain the following set of equations :  \hspace{3cm} \linebreak
\hspace{3cm} \parbox{3cm}{\begin{equation} \label{eq_dcstrich}
\Dg{c}' = - x^2 \tilde {V}_c \end{equation}}
\hspace{2cm} \parbox{4cm}{\begin{equation} \label{eq_vcstrich} \tilde V_c' = -2 \tilde V_c + \Phiflex \end{equation}}
\parbox{4cm}{\begin{equation} \label{eq_dgammastrich}
\Delta_{\gamma}' = -\frac{4}{3} x^2 \tilde V_{\gamma} \end{equation}} \linebreak

\parbox{5.5cm}{\begin{equation} \label{grand_equation_aa} \tilde V_{\gamma}' =  \frac{1}{4} \Delta_{\gamma}- \tilde V_{\gamma}
 + \Omega_{\nu} \tilde{\Pi}_{\nu} + 2\Phiflex \end{equation}}  
\parbox{3cm}{\begin{equation} \Dg{b}' = -x^2 \tilde V_{\gamma} \label{eq_dgbaryonstrich} \end{equation}} 
\parbox{4cm}{\begin{equation} \Delta_{\nu}' = -\frac{4}{3} x^2 \tilde V_{\nu} \label{eq_dnustrich} \end{equation}} \linebreak

\parbox{7cm}{\begin{equation} \label{grand_equation_bb} \tilde V_{\nu}' =  \frac{1}{4} \Delta_{\nu} - \tilde V_{\nu} 
-\frac{1}{6}x^2 \tilde \Pi_{\nu} + \Omega_{\nu} \tilde{\Pi}_{\nu} + 2\Phiflex \end{equation}} 
\parbox{4cm}{\begin{equation} \tilde \Pi_{\nu}'= \frac{8}{5}\tilde V_{\nu}-2 \tilde \Pi_{\nu} \end{equation}} \linebreak

\begin{equation} \label{eq_q1} \Deltaphi' = 3 (\wq-1)\bigg[  \Deltaphi + 3(1+\wq)\left\{ \Phiflex + \Omega_{\nu} 
\tilde{\Pi}_{\nu}\right \} + \left \{3 - \frac{x^2}{3(\wq -1)} \right \}\, (1+\wq) \Vphitilde \bigg ] \end{equation}

\parbox{6.1cm}{\begin{equation}
\label{eq_q2}
\Vphitilde'= 3\Omega_{\nu} \tilde{\Pi}_{\nu}  +\frac{\Deltaphi}{1+\wphi} + \Vphitilde   +4 \Phiflex 
\end{equation}} 
\parbox{9.1cm}{\begin{equation} \label{eq_poisson_sum}
\Phiflex=- \frac{ \sum\limits_{\alpha =c,b,\gamma,\nu,q} \Omega_{\alpha} ( \Delta_{\alpha} +3(1+w_{\alpha})
\tilde{V}_{\alpha})}{ \sum\limits_{\alpha =c,b,\gamma,\nu,q} 
 3(1+w_{\alpha})\Omega_{\alpha} + \frac{2 x^{2}}{3}} -\Omega_{\nu} \tilde{\Pi}_{\nu}
\end{equation}}    \linebreak
with the gauge-invariant Newtonian potential $\Phiflex$.
We denote the derivative $\textrm{d}/\textrm{d} \ln x$ with a prime.
The gauge-invariant energy density contrasts  $\Delta_{\alpha}$, the velocities
$\tilde{V}_{\alpha}$ and the shear $\tilde \Pi_{\nu}$ are the ones
found in the literature \cite{Bardeen:kt,Kodama:bj,Durrer:2001gq}, except that 
we factor out powers of $x$
from the velocity and shear defining $\tilde{V} \equiv V/x$ and $\tilde{\Pi}_{\nu} \equiv x^{-2} \Pi_{\nu}$.
The index $\alpha$ runs over the five
species in our equations, quintessence is assigned the subscript $q$. We assume tight
coupling between photons and baryons. The equation of state
$w=\bar{p}/\bar{\rho}$ takes on the values $w_{c}=w_b=0$,
$w_{\gamma}=w_{\nu}=1/3$ and $\wphi$ is left as a free parameter.



We define the perturbation vector as
\parbox{9.5cm}{\begin{equation}
\bm{U}^{T} \equiv (\Dg{c},\,\tilde{V}_{c},\,\Delta_{\gamma},\,\tilde{V}_{\gamma},
\,\Dg{b},\,\Delta_{\nu},\,\tilde{V}_{\nu},\,\tilde{\Pi}_{\nu},\,\Deltaphi,\,\Vphitilde).
\end{equation}}
The matrix $A(x)$ can easily be read off from equations (\ref{eq_dcstrich})-(\ref{eq_q2}).
This enables us to discuss the problem of 
specifying initial conditions in a systematic way. 

The initial conditions are specified for modes well outside the horizon, i.e.
$x \ll 1$.  In this case,  the r.h.s. of equations
(\ref{eq_dcstrich}),\ (\ref{eq_dgammastrich}), (\ref{eq_dgbaryonstrich}) and (\ref{eq_dnustrich})
can be neglected, provided $\tilde{V}_{\alpha}$ does not diverge $
\propto x^{-2}$ or faster for $x^2\to0$. 

The general solution to Equation (\ref{eqn::matrix}) in the (ideal) case of 
a truly constant $A$ would be 
\ee\label{eigenvector_decomposition}
\bm{U}(x) = \sum\limits_i c_{i} \left( \frac{x}{x_0}\right)^{\lambda_i}\bm{U}^{(i)},
\eee
where $\bm{U}^{(i)}$ are the eigenvectors of $A$ with eigenvalue $\lambda_i$ and the
time independent coefficients $c_i$ specify the initial contribution of $\bm{U}^{(i)}$ towards a 
general perturbation $\bm{U}$.
As time progresses, components corresponding to the largest eigenvalues $\lambda_i$
will dominate. Compared to these ``dominant'' modes, initial contributions  
in the direction of eigenvectors  $\bm{U}^{(i)}$ with smaller ${\rm Re}(\lambda_i)$ 
decay.
In our case, the characteristic polynomial of $A(x)$ indeed has a fourfold degenerate
eigenvalue $\lambda = 0$ in the limit $x^2 \to 0$, the other six remaining eigenvalues are negative.
We therefore need to solve $A(x) \bf{U} =0$ which is equivalent to setting 
the l.h.s. of Equations (\ref{eq_dcstrich})-(\ref{eq_q2}) 
equal to zero and using $\omca=\omba=x^2=0$.
Then Equations (\ref{eq_dcstrich}),
(\ref{eq_dgammastrich}), (\ref{eq_dgbaryonstrich})  and (\ref{eq_dnustrich}) are automatically
satisfied (provided $\tilde{V}_{\alpha}$ does not diverge 
$\propto x^{-2}$ or faster), and  Equations (\ref{eq_vcstrich}),(\ref{grand_equation_aa}),(\ref{grand_equation_bb})-(\ref{eq_q2}) 
yield non-trivial constraints for the components of $\bm{U}$: \hspace{8cm}  \linebreak

\parbox{4.5cm}{\begin{equation} \label{constraint_equation_start}
2 \tilde V_c-\Phiflex = 0 \end{equation} } \hspace{1cm} \parbox{6.0cm}{ \begin{equation} 1/4 \Delta_{\gamma} - \tilde V_{\gamma}+\Omega_{\nu}\tilde \Pi_{\nu}+2\Phiflex = 0 \end{equation} }  \linebreak

\parbox{6.5cm}{\begin{equation}
1/4 \Delta_{\nu} - \tilde V_{\nu}+ \Omega_{\nu}\tilde \Pi_{\nu} +2\Phiflex = 0 \end{equation}} \hspace{0.1cm} \parbox{6cm}{\begin{equation} 8/5 \tilde V_{\nu}- 2\tilde \Pi_{\nu} =  0   \end{equation}} \linebreak

\parbox{7cm}{\begin{equation} \label{constraint_equation_onebeforeend}  3 \Omega_{\nu}\tilde \Pi_{\nu} +\Deltaphi /(1+\wphi)+ 3 \Vphitilde + 3 \Phiflex =  0 \end{equation} } \hspace{0.1cm} \parbox{7cm}{\begin{equation} \label{constraint_equation_end}
3\Omega_{\nu}\tilde \Pi_{\nu}+\Deltaphi/(1+\wphi)+ \Vphitilde +4\Phiflex= 0 
\end{equation}} \linebreak



Following the existing literature, we use the 
gauge-invariant entropy perturbation \cite{Kodama:bj} between two species $\alpha$ and $\beta$, 
as well as the  gauge-invariant curvature perturbation on
hyper-surfaces of uniform energy density of species $\alpha$ in order to classify the physical modes\cite{Bardeen:qw,bardeen_isocurvature_book,Lyth:2002my,Wands:2000dp} :  

\parbox{6cm}{\begin{equation}
S_{\alpha:\beta}= \frac{\Delta_{\alpha}}{1+w_{\alpha}}-\frac{\Delta_{\beta}}{1+w_{\beta}},
\end{equation}}
\parbox{7.5cm}{\begin{equation}
\zeta_{\alpha}= \left(H_L+\frac{1}{3} H_T\right)+\frac{\delta \rho_{\alpha} }{3(1+w_{\alpha})\bar \rho_{\alpha}}.
\end{equation}}  \linebreak

In our variables, these expressions take on the manifestly gauge-invariant form
{\begin{equation}
\zeta_{\alpha}=\frac{\Delta_{\alpha}}{3(1+w_{\alpha})} \ , \ \ \  \zeta=\frac{\sum_{\alpha} \Delta_{\alpha} \Omega_{\alpha}}{\sum_{\alpha} 3(1+w_{\alpha})\Omega_{\alpha}}.
\end{equation}



The first (rather intuitive) perturbations one would try to find are adiabatic
perturbations, which are specified by the adiabaticity conditions
$S_{\alpha:\beta}=0$ for all pairs of components, i.e.
\begin{equation}\label{adiabatic_conditions}
\Delta_{\nu}=\Delta_{\gamma}=\frac{4}{3} \Dg{c} = \frac{4}{3} \Dg{b},
\end{equation}
Using the six constraint Equations (\ref{constraint_equation_start})-(\ref{constraint_equation_end}), we obtain the adiabatic
mode. (Due to limited lenght of this article we cannot quote the full results, we therefore refer the reader to Doran et. al. \cite{Doran:2003xq} for more details.) We also conclude that quintessence is automatically adiabatic if CDM, baryons,
neutrinos and radiation are adiabatic, independent of the quintessence
model for  as long as we are in the tracking regime.

Let us next consider the  neutrino isocurvature mode. For this, we
require that CDM, baryons  and radiation are adiabatic, while $S_{\nu:\gamma} \neq 0$
and that the gauge-invariant curvature perturbation vanishes:
\begin{equation}
\zeta = 0, \ \ \ \Dg{c}= \Dg{b} = \frac{3}{4} \Delta_{\gamma}.
\end{equation}
Using this and Equations (\ref{constraint_equation_start})-(\ref{constraint_equation_end}) leads to 
the neutrino-isocurvature perturbation.

It is important to note that we did not require quintessence to be
adiabatic. One can see from the neutrino isocurvature vector that
$\Deltaphi=0$, and as a consequence quintessence is not adiabatic with
respect to either neutrinos, radiation, baryons or CDM. 
Hence, we could just as well have labeled this vector ``quintessence
isocurvature''. 

The CDM isocurvature mode is characterized by $S_{c:\gamma} \neq 0$, $\zeta=0$
and adiabaticity between photons, neutrinos and baryons. Similarly, for the baryon isocurvature 
mode we require $S_{b:\gamma} \neq 0$, $\zeta=0$.

The adiabatic, \isocdm, \isobar\ and neutrino isocurvature-
vector are linearly independent. We have therefore identified 
four modes corresponding to the fourfold degenerate eigenvalue zero of $A(x)$.
These four vectors span the subspace of dominant modes in the
super-horizon limit. Arbitrary initial
perturbations may therefore be represented by projecting a
perturbation vector $\bm{U}$ at initial time into the
subspace spanned by the four aforementioned vectors, as this is the
part of the initial perturbations which will dominate as time progresses.



\begin{figure}[ht]
   \begin{center}
    \includegraphics[scale=0.37,angle=-90]{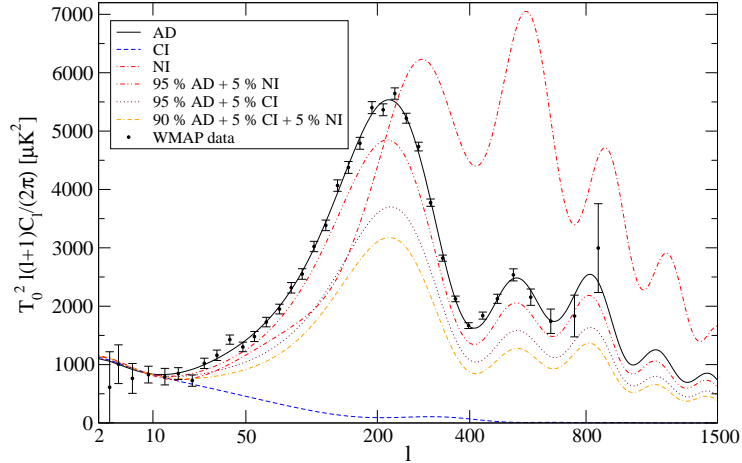}
   \caption{CMB Temperature spectra as a function of multipole $l$ in an early quintessence cosmology. 
      The pure adiabatic (AD), CDM isocurvature (CI), neutrino isocurvature (NI) mode and
      three different combinations of these dominant modes are plotted. For comparison with experimental data 
      we also give the WMAP measurements of the CMB \protect\cite{Spergel:2003cb}. The spectrum of the pure \isobar\ mode is 
      essentially identical to that of the pure \isocdm\ mode. All spectra have been normalized to the same power at $l=10$}
    \label{example}
  \end{center}
\end{figure}



We use a modified version
of {\sc cmbeasy} \cite{Doran:2003sy,Seljak:1996is} to compute CMB spectra corresponding
to different initial conditions for an early quintessence cosmology
with parameters as in model A of \cite{Caldwell:2003vp}. We set the
spectral index of the isocurvature modes identical to the spectral
index of the pure adiabatic mode, $n_s = 0.99$. Comparison with the WMAP
data in the same figure shows that non-adiabatic initial perturbations
are strongly constrained.


\subsection*{Conclusion} \label{section_conclusion}
We have investigated perturbations in a radiation-dominated universe containing
quintessence, CDM, neutrinos, radiation and baryons in the tight
coupling limit. The perturbation evolution has been expressed as 
a differential equation involving a matrix acting on a vector comprised
of the perturbation variables.  This formulation leads to a systematic
determination of the initial conditions. In particular, we find that due to
the presence of tracking scalar quintessence no additional dominant mode is introduced. 
This fact is beautifully transparent in the matrix language. In total, we find four dominant modes and choose them as 
adiabatic, \isocdm, \isobar\ and \isonu. For the \isonu\ mode, quintessence
automatically is forced to non-adiabaticity. Hence, we could have as well labeled
the \isonu\ mode as quintessence isocurvature.
To demonstrate the influence on the cosmic microwave background anisotropy
spectrum, we have calculated spectra for all modes. A detailed study may provide ways to
put additional constraints on quintessence models or tell us more
about the initial perturbations after inflation.


\bibliographystyle{unsrt}

\end{document}